# Magnetoplasmonic Quasicrystals

**Andrey N. Kalish,[1,2] Roman S. Komarov,[2] Mikhail A. Kozhaev,[1,3] Venu Gopal Achanta,[4] Sarkis A. Dagesyan,[2] Alexander N. Shaposhnikov,[5] Anatoly R. Prokopov,[5] Vladimir N. Berzhansky,[5] Anatoly K. Zvezdin,[1,3] Vladimir I. Belotelov,[1,2,*]**

[1]*Russian Quantum Center, Skolkovskoe shosse 45, Moscow 121353, Russia*
[2]*Faculty of Physics, Lomonosov Moscow State University, Leninskie Gory, Moscow 119991, Russia*
[3]*Prokhorov General Physics Institute of the Russian Academy of Sciences, Vavilov str. 38, Moscow 119991, Russia*
[4]*Tata Institute of Fundamental Research, Homi Bhabha Road, Navy Nagar, Colaba, Mumbai 400005, India*
[5]*Vernadsky Crimean Federal University, Vernadsky Ave. 4, Simferopol 295007, Russia*
**Corresponding author: belotelov@physics.msu.ru*

**Nanostructured magneto-optical materials sustaining optical resonances open very efficient way for light control via magnetic field, which is of prime importance for telecommunication and sensing applications. However, usually their response is narrowband due to its resonance character. Here we demonstrate and investigate a novel type of the magnetoplasmonic structure, the magnetoplasmonic quasicrystal, which demonstrates unique magneto-optical response. It consists of the magnetic dielectric film covered by a thin gold layer perforated by slits forming a Fibonacci-like binary sequence. The transverse magneto-optical Kerr effect (TMOKE) acquires controllable multiple plasmon-related resonances resulting in a magneto-optical response in a wide frequency range. The broadband TMOKE is valuable for numerous nanophotonics applications including optical sensing, control of light, all-optical control of magnetization etc. On the other hand, TMOKE spectroscopy is an efficient tool for investigation of the peculiarities of plasmonic quasicrystals.**

## 1. INTRODUCTION

Since 1970s quasicrystalline structures have attracted attention of researchers. They were experimentally discovered by Shechtman et al. [1], for which he was awarded the Nobel Prize in Chemistry in 2011. Quasicrystals are non-periodic but ordered structures [2]. Contrary to the periodic crystalline structures, they do not possess translational symmetry but long-range ordering is present. For periodic structures the reciprocal lattice is given by equidistant set of reciprocal vectors. Thus for one-dimensional periodicity of period d the reciprocal vectors **G** are given by $\mathbf{G}=(2\pi/d\ m)\mathbf{e}$, where **e** is the unit vector along the periodicity direction and m is an integer. Unlike periodic structures, the quasiperiodic ones are characterized by discrete but non-equidistant reciprocal lattice and corresponding unusual diffraction patterns. The example of one-dimensional quasicrystalline pattern is the Fibonacci binary sequence [3]. It is constructed by the infinite specific sequence of '0' and '1' numbers.

Two-dimensional quasicrystalline structures can possess rotational symmetry $C_n$ of the orders prohibited for the periodic structures, such as $n = 5, 7$ and higher orders. An example of that is Penrose tiling that possesses $C_5$ rotational symmetry [4]. Quasicrystalline substances are usually metallic alloys and they provide high hardness, low friction coefficient, low heat conductivity and wide optical absorption bands. Among potential applications of quasicrystals there are two-phase materials containing steel with high strength and plasticity at high temperatures, coatings, for example, for frying pans, and solar energy absorbers and reflectors [2].

The concept of quasicrystals has been implemented in plasmonics as well [4-12]. Study of plasmonic structures has been of prime interest for the last decade regarding control of plasmonic modes and near-field manipulation [13,14], design of optical properties [15], high sensitivity to surrounding media [16,17], all-optical light control [18-20], and ultrafast magnetism [21,22]. Plasmonic quasicrystals in the form of metal-dielectric nanostructures with quasicrystal pattern that supports excitation of surface plasmon-polaritons (SPP) were designed. Due to discrete non-equidistant reciprocal space and rotational symmetry of quasicrystals, larger number of resonances associated with SPP modes appear and therefore such structures demonstrate a broadband optical response. Moreover, it is polarization-independent for the 2D structures [7]. The advantage of quasicrystalline structures over periodic and non-periodic ones is that they possess rich and designable reciprocal lattice that governs the optical diffraction and dispersion of the eigenmodes, and therefore offers designable broadband optical response [23]. It comes from the fact that the reciprocal lattice is strongly dependent on geometrical parameters of the structure, while for the periodic structures it is defined solely by the period.

However, the potential of quasicrystalline structures for magneto-optics has not been revealed yet. At the same time, magneto-optics is a powerful tool for manipulation and probing of optical properties of different materials and structures. In detail, vast research has been carried out during recent decade on the enhancement of magneto-optical effects in plasmonic structures [24-41]. Combining magneto-optics with plasmonics provides significant advance in magneto-optical properties. It was demonstrated that in systems containing bismuth iron-garnet films [42,43] conventional magneto-optical effects such as the transverse Kerr effect (TMOKE) are resonantly enhanced in magnetoplasmonic crystals [37,38], and also novel promising effects arise [39]. However, these effects are of resonant nature, related to the excitation of eigenmodes of the structure, such as cavity modes, SPPs, waveguide and quasi-waveguide modes etc. that leads to narrow spectral range of the magneto-optical response.

In the present work, we propose and demonstrate an approach for forming a broadband magneto-optical response using one-dimensional magnetoplasmonic quasicrystals. We design and fabricate the magnetoplasmonic quasicrystal and measure its optical transmittance and the TMOKE and compare them to the case of periodic structure. We find that the TMOKE in a quasicrystalline structure has multiple-resonance character and, moreover, it is a sensitive tool for investigation of the quasicrystal spectral properties.

## 2. MAGNETOPLASMONIC QUASICRYSTAL SAMPLES AND EXPERIMENTAL SET-UP

The considered 1D magnetoplasmonic quasicrystalline structure is formed by a metallic quasicrystal grating on top of the smooth magnetic dielectric layer on a substrate (Fig. 1(a)). The sequence of metal stripes and air slits of the grating can be described by symbols ‘1’ and ‘0’. Our structure is based on the 1D binary Fibonacci sequence, where ‘0’ is substituted by ‘010’. The final formula of the structure is thus the following:

‘101011010101011010110101010110101010110101101010101010110
1011010101011010101011010110101010110101010110101101010
10101101011010101011010101011010110101010110101101011…‘ (1)

Such structure was chosen to achieve rather large spectral density of resonances.

The schematic of a part of the structure corresponding to the underlined blocks in Eq. (1) is shown in Fig. 1(a). For comparison we also consider a periodic crystal that is described by sequence ‘10101010101…’ (Fig. 1(b)).

The metal grating of the experimentally studied samples was fabricated by electron beam lithography of the thermally deposited 80-nm-thick gold layer. The air slit width corresponding to single ‘0’ in the binary sequence is 80 nm and the metal stripe width corresponding to single ‘1’ in the sequence is 600 nm. The magnetic dielectric is a bismuth substituted iron-garnet film of composition $Bi_{1.5}Gd_{1.5}Fe_{4.5}Al_{0.5}O_{12}$ [46]. It was fabricated by reactive ion beam sputtering on (111) gadolinium gallium garnet substrate. The thickness of the magnetic film was made rather small, 80 nm, to exclude the waveguide mode excitation in the considered frequency range. Thus, only SPPs can be excited.

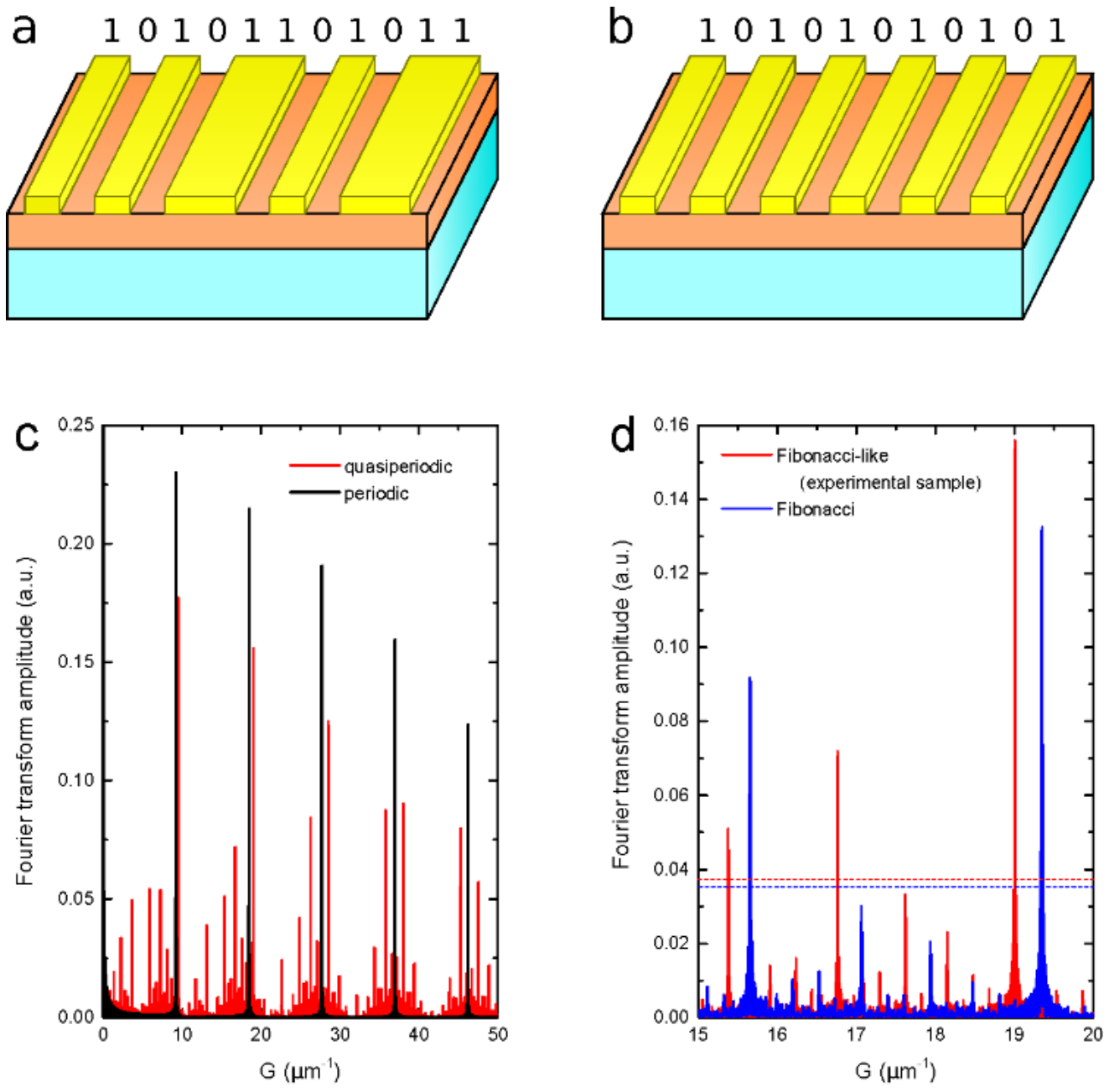

Fig. 1. (a,b) Schematics of the experimentally studied one-dimensional magnetoplasmonic quasicrystal (a) and the magnetoplasmonic crystal (b). Golden stripes are on top of the magnetic dielectric film deposited on a substrate. (c) The Fourier transform of the experimentally studied quasicrystal formed by Fibonacci-like sequence (red line) in comparison with the spectrum of the periodic crystal having the same width of gold stripes as the quasicrystal (black line). (d) The Fourier transform of the experimentally studied quasicrystal formed by the Fibonacci-like sequence (red line) in comparison with the spectrum of the quasicrystal formed by the pure Fibonacci sequence (blue line). Dashed horizontal lines indicate 25% level with respect to the highest peaks.

Optical and magneto-optical spectra of the plasmonic quasicrystal samples were measured by the following experimental set-up. Tungsten halogen lamp is used as a light source. After the lamp light passes through the fiber in order to obtain homogeneous point-like light source, it is collimated with achromatic 75mm lens and focused onto the sample with another achromatic 35mm lens to a spot of about 200 μm in diameter. To perform the TMOKE measurements the sample is placed in uniform external magnetic field of 200 mT generated by electromagnet in the direction along the gold stripes. After the sample, the light is collimated with 20x microobjective and detected with the spectrometer. The spectrometer slit was oriented perpendicularly to the gold stripes, so only light with perpendicular to gold stripes incidence plane was detected. Spectral and angular decomposed light signal is detected by the spectrometer with CCD camera.

## 3. FOURIER SPECTRA OF THE PLASMONIC STRUCTURES

Spectrum of the SPPs excited by the incident light in a plasmonic grating structure is determined by the reciprocal lattice vectors **G** which enter the phase matching condition:

$$\boldsymbol{\beta} = \mathbf{k}_{\parallel} + \mathbf{G}\,, \quad (2)$$

where $\boldsymbol{\beta}$ is the SPP propagation vector, and $\mathbf{k}_{\parallel}$ is the in-plane component of the incident light wave vector. The reciprocal lattice vectors **G** are found from the Fourier transform of the metal pattern as the coordinates in the reciprocal space of the Fourier transform peaks.

Fig. 1(c) depicts calculated absolute values of the Fourier transforms of the quasicrystal (red line) and the periodic crystal (black line)

patterns of the considered samples. The lowest peaks in the quasicrystal Fourier transform with amplitudes smaller than 0.01 a.u. can be neglected as they appear as a numerical calculation error. Moreover, low peaks do not contribute significantly to the excitation of eigenmodes, as the corresponding excitation efficiency is small. The reciprocal lattice for the quasicrystal is discrete and it is far denser compared to that of the periodic crystal. The discrete non-equidistant reciprocal lattice confirms that the structure is non-periodic but has some long-range order. The reciprocal vectors for the periodic crystal form a discrete equidistant lattice with their values $G_m^{(per)}=m\cdot 2\pi/d$, where $d$ is the period and $m$ is an integer. For the considered parameters of the grating $G_m^{(per)}=m\cdot 9.24$ μm$^{-1}$. Due to equality of the width of the elements for both structures the reciprocal lattices correlate with each other. Namely, the reciprocal vectors of the quasicrystal lattice form bands that are located in the vicinities of $G_m^{(per)}$.

Reciprocal lattices of quasicrystals can be tuned by adjusting their type of sequence. Fig. 1(c,d) shows Fourier spectra both for the experimentally considered structure (Eq. (1), red lines in (c) and (d)) and for the pure Fibonacci sequence (blue line in (d)). Spectral positions of the peaks and their amplitudes are different. In detail, in the range of the **G** vectors from 15 μm$^{-1}$ to 20 μm$^{-1}$ which produces resonances in the studied wavelength range from 750 nm to 1000 nm, the experimental sample has 3 peaks higher than 25%-level of the main peak. At the same time, the pattern based on the pure Fibonacci sequence provides only two such peaks (Fig. 1(d)). Another advantage of the experimentally considered structure is that the highest peaks are located closer to each other, so their spectral density is larger than the one for the Fibonacci structure. Consequently, the quasicrystals formed by different types of sequences have quite different Fourier transform and therefore the SPP spectra.

## 4. TRANSMISSION SPECTRA OF PLASMONIC QUASICRYSTALS

Transmission spectra of the plasmonic samples demonstrate some resonant features which are dispersive with respect to the angle of incidence (Fig. 2(a,b)). They are due to the SPPs propagating at the interface between the magnetic film and gold grating as is confirmed by the calculated dispersion of the SPP resonances (green dashed lines). The dispersion was calculated by Eq. (2) with propagation constant $\beta_0$ taken from approximation of a smooth single-interface between two semi-infinite metal and dielectric media: $\beta_0=2\pi/\lambda\ [(\varepsilon_m\ \varepsilon_d)/(\varepsilon_m+\varepsilon_d)]^{1/2}$, where $\varepsilon_m$ and $\varepsilon_d$ are dielectric constants of metal and dielectric, respectively, and $\lambda$ is the vacuum wavelength. The dielectric constants were taken from [44] for gold and from [42] for magnetic dielectric. To take into account the finiteness of the magnetic film thickness, the presence of slits and the difference in magnetic chemical composition with respect to [42] $\varepsilon_d$ was varied to obtain best fit with experimentally observed spectral positions of optical resonances in both samples. The best correspondence was achieved when $\varepsilon_d$ was multiplied by 0.95 for the plasmonic crystal and by 0.91 for the plasmonic quasicrystal compared to values from [42]. The difference in multiplicators comes from the difference in grating patterns, as the periodic and quasiperiodic crystals have different effective filling factors of metal.

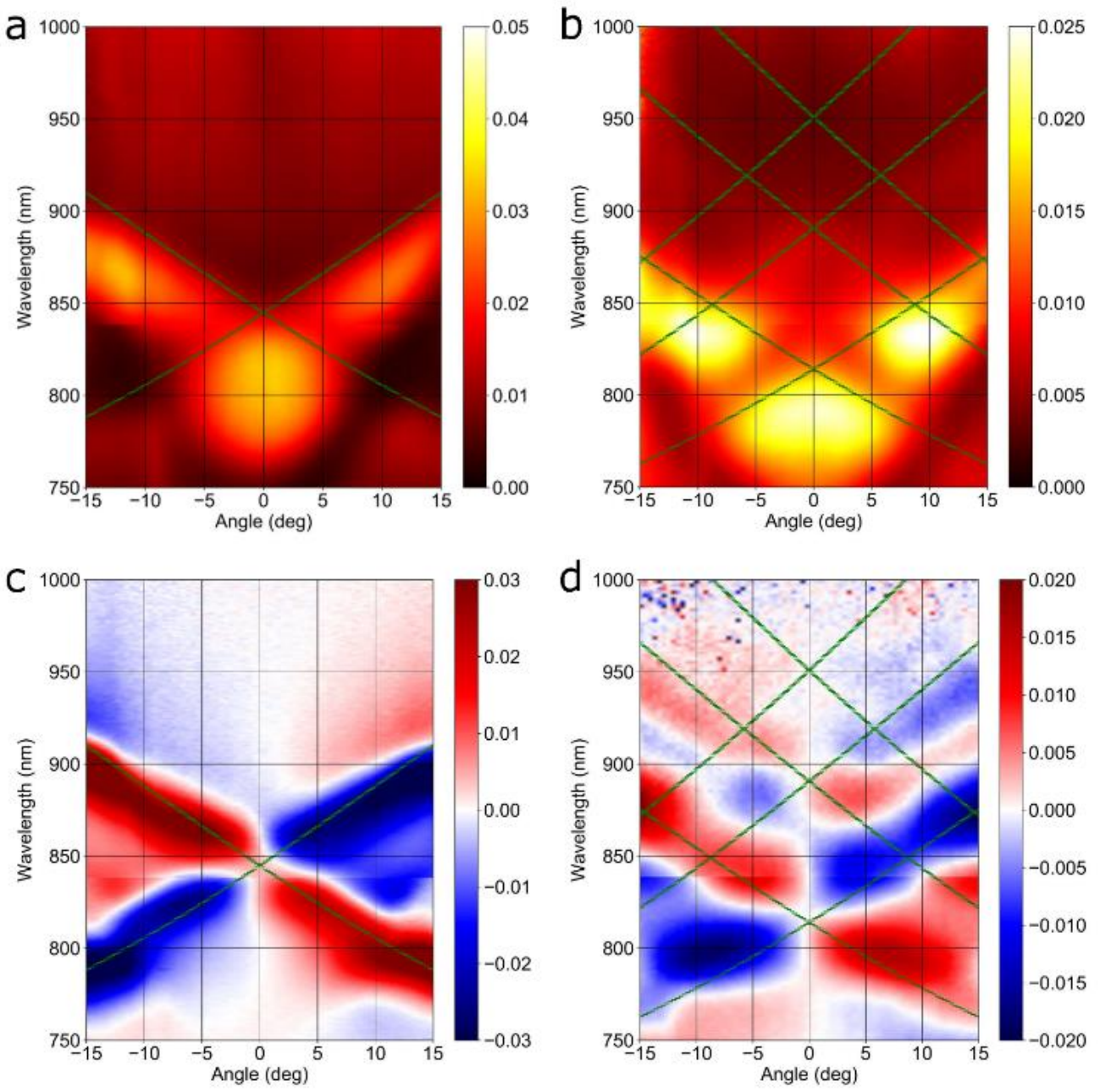

Fig. 2. The dependence of transmittance (a,b) and the TMOKE (c,d) on the wavelength and the incident angle for the magnetoplasmonic crystal (a,c) and quasicrystal (b,d). Green lines show the calculated dispersion curves for the surface plasmon polaritons.

Excitation of SPPs leads to Fano-shape resonances in transmittance. In the experimentally studied samples, main transmittance resonances are due to SPPs excited by the second diffraction order with $G$=18.48 μm$^{-1}$ and $G$=19.01 μm$^{-1}$, in the periodic and quasiperiodic structures, respectively (Fig. 1(c)). For the normal incidence it corresponds to the transmittance peaks at $\lambda$=820 nm (for the crystal, Fig. 2(a)) and $\lambda$=790 nm (for the quasicrystal, Fig. 2(b)). For the oblique incidence SPP resonances are split into low-frequency and high-frequency ones and demonstrate dispersion versus angle of incidence which is seen in Fig. 2(a,b) by peaks and dips following SPP dispersion curve (green lines). Though the dispersion lines for the plasmonic quasicrystal predict more resonances in the range from $\lambda$=880 nm to $\lambda$=950 nm they are hardly seen in transmittance. As a result, transmittance spectra for the crystal and its quasiperiodic counterpart look rather similar.

## 5. REVEALING PLASMONIC MODES IN QUASICRYSTALS FROM TMOKE SPECTRA

Since magneto-optical effects in plasmonic structures are enhanced by the excitation of eigenmodes they can be used for investigation of the quasicrystals spectra. The impact of the magnetization **M** on the SPP dispersion is given by [45]:

$$\boldsymbol{\beta} = \boldsymbol{\beta}_0\left(1+\alpha\left(\boldsymbol{\beta}_0\cdot[\mathbf{M}\times\mathbf{n}]\right)\right), \qquad (3)$$

where **n** is the surface normal, $\boldsymbol{\beta}_0$ is the mode propagation vector for zero magnetization, and $\alpha$ is the proportionality coefficient dependent on permittivities of dielectric and metal media. Here we keep only linear in M contribution. On the other hand, the TMOKE, if observed in transmitted light, compares intensity of the transmitted light $T$ for the opposite directions of the transversal magnetization $T(\mathbf{M})$ and $T(-\mathbf{M})$ and is defined by $\delta=[T(\mathbf{M})-T(-\mathbf{M})]/T(0)$, where $T(0)$ is transmittance through the non-magnetized sample. The magnetically induced change in dispersion gives rise to the enhancement of the TMOKE at the slopes of the plasmonic resonances in optical spectra. It happens because the spectral positions of plasmonic resonances are shifted by the transverse magnetization and therefore, at their slopes, where the absolute value of the transmittance derivative with respect to the wavelength reaches its maximum, large difference between transmittance for oppositely magnetized cases appears. According to Eq. (2), large density of the reciprocal lattice vectors produces multiplicity of plasmonic resonances and, therefore, of the TMOKE resonances.

Experimentally measured TMOKE spectra are shown in Fig. 2(c,d). In contrast to the transmission spectra, they are quite different for the crystal and quasicrystal samples. It highlights high sensitivity of the TMOKE to the excitation of the eigenmodes of the structure. Let us consider them in detail. For the periodic crystal, the TMOKE spectrum demonstrates pronounced resonances at the slopes of the transmittance peaks and vanishes at normal incidence due to symmetry reasons (Fig. 2(c)). At the oblique incidence, two pronounced TMOKE peaks of opposite signs are observed at around $\lambda$=850 nm. They are related to the low-frequency and high-frequency branches of the plasmonic dispersion and their opposite sings are due to the propagation of SPPs in forward and backward directions. The TMOKE reaches 2% at the extrema of the spectra. No noticeable TMOKE response is observed for $\lambda$>900 nm.

However, for the plasmonic quasicrystal the TMOKE demonstrates much richer response: three pairs of resonances appear. Apart from the first pair at around $\lambda$=820 nm (for small incidence angles) that is quite similar to the resonances for the periodic structure at $\lambda$=850 nm, two other pairs appear at around $\lambda$=890 nm and $\lambda$=950 nm corresponding to $G$=16.76 μm$^{-1}$ and $G$=15.39 μm$^{-1}$ (see Fig. 1(c)). The resonance at $\lambda$=890 nm is well pronounced and provides the TMOKE as large as 0.8%. The last one is rather weak and the TMOKE doesn't exceed 0.3% though is still detectable above the noise level. For increasing angle of incidence the pairs of resonances split further and mix with their neighbours.

Fourier spectrum of the quasicrystal indicates that there are two more reciprocal vectors of $G$=17.62 μm$^{-1}$ and $G$=18.15 μm$^{-1}$ in the vicinity of $G$=18.48 μm$^{-1}$ but they have relatively low Fourier amplitudes (0.03 and 0.02, respectively) and, therefore, do not produce any notable resonances even in TMOKE.

Thus, the plasmonic quasicrystals offer multiplicity of the plasmonically mediated TMOKE. Fig. 3(a) demonstrates it more clearly. The TMOKE spectrum for the range of wavelengths from $\lambda$=750 to $\lambda$=1000 nm is shown for the incident angle of 2 degrees for the periodic and quasicrystalline structures. For the former only one pair of resonances (marked by 'P') is observed while, the latter clearly exhibits two close pairs of resonances ('Q1' and 'Q2') and also the third weaker one ('Q3'). It confirms that plasmonic quasicrystals allow much broader magneto-optical response.

As we saw in Fig. 1(c,d) the quasicrystals formed by different type of sequences have quite different Fourier transform and therefore SPP spectra. However, most of the resonances do not have large efficiency and are hardly observable in transmission or reflection. At the same time, measuring TMOKE allows to reveal these resonances and identify different quasicrystal patterns.

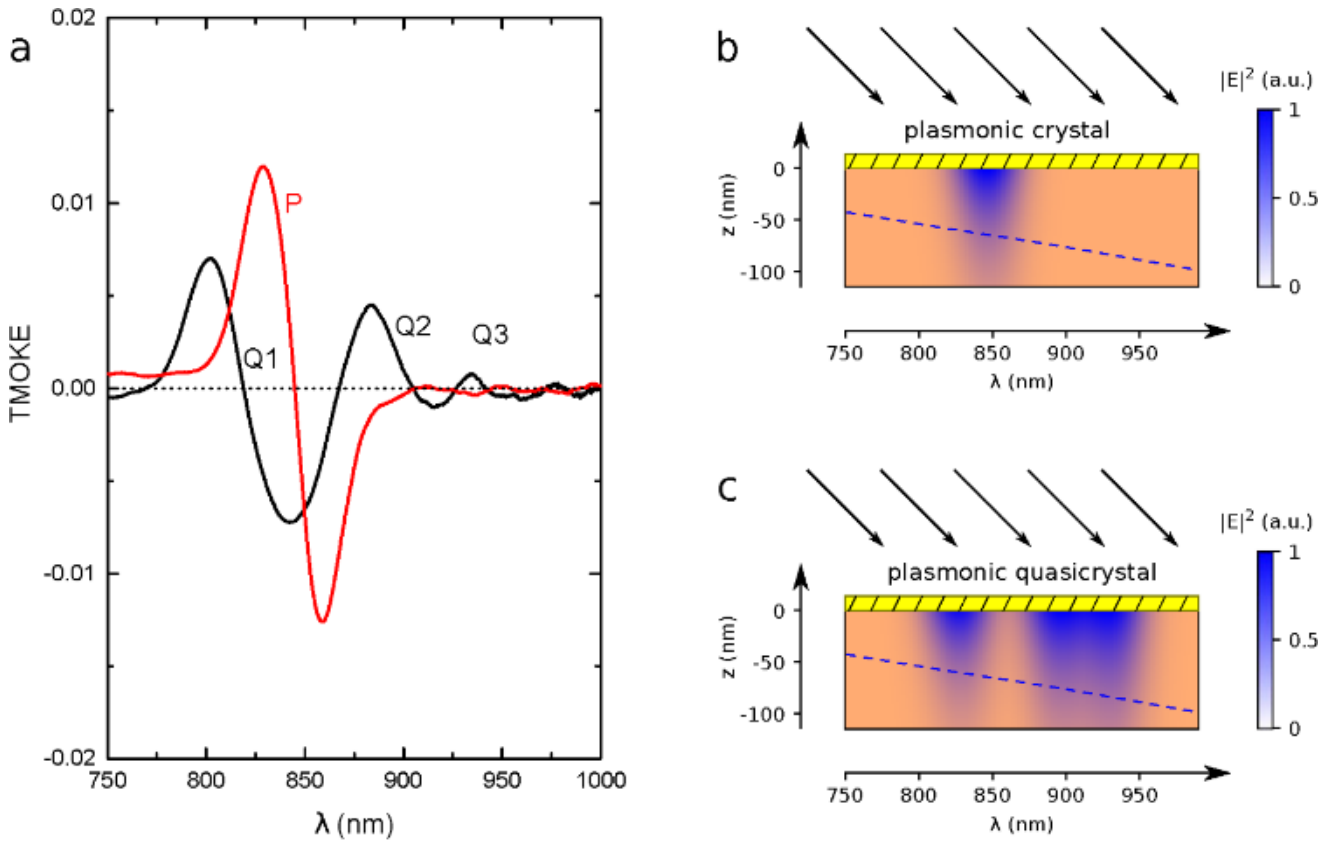

Fig. 3. (a) The TMOKE spectrum for the incident angle of 2°. (b,c) The schematically shown dependence of the SPP penetration depth on the wavelength and the near-field of the excited SPPs for the magnetoplasmonic crystal (b) and quasicrystal (c). Intensity of blue color indicates the SPP amplitude versus wavelength (horizontal axis) and depth (vertical axis). Blue dashed line indicates depth in the magnetic dielectric film at which the amplitude of SPP decreases by *e* times.

The multiplicity of the excited plasmonic modes attracts attention also because they possess different values of the penetration depth (Fig. 3). Estimations show that for the plasmonic resonances shown in Fig. 3(a), the SPP penetration depth in the magnetic dielectric varies approximately from 70 to 100 nm. This fact opens new possibilities for manipulation of the optical near field, 3D sensing, control of the inverse magneto-optical effects and optically-induced magnetization.

After demonstration of the concept of magnetoplasmonic quasicrystals the next step is the design of plasmonic quasicrystalline structures in order to obtain desired optical response. Quasicrystalline structures provide designable reciprocal lattice, i.e. the set of reciprocal vectors and therefore, dispersion of eigenmodes, by means of adjusting geometrical parameters. In particular, plasmonic quasicrystals offer designable spectrum of magneto-optical response for light modulation, which is prosperous for parallel light information processing at several frequencies.

Furthermore, the plasmonic quasicrystals are promising for achieving other broadband magneto-optical effects related to the excitation of eigenmodes. If the structure supports waveguide modes then there are many resonances for TE and TM modes with the resonant wavelengths close to each other. This condition is favorable for the enhancement of the Faraday effect, as the TE-TM conversion is the most effective. Besides that, the magnetization affects the waveguide mode polarization that usually leads to the longitudinal magnetophotonic effect (LMPIE) in plasmonic structures. The dense dispersion of the waveguide modes might lead to multiple LMPIE resonances.

## 6. CONCLUSION

We have proposed and demonstrated a novel structure - magnetoplasmonic quasicrystal for getting significant magneto-optical effects in the broadband wavelength range. It is based on magnetic dielectric film and the gold film perforated with the subwavelength slits forming Fibonacci-like binary sequence. While transmission spectra of the periodic and quasiperiodic patterns are quite similar, the TMOKE spectra demonstrate significant difference. Namely, for the quasicrystal the magneto-optical response is much more abundant. It shows that TMOKE spectroscopy is an efficient tool for investigation of the peculiarities of plasmonic quasicrystals.

Due to larger density of the discrete peaks in the Fourier transform of the quasicrystal the additional plasmonic resonances appear leading to multiple TMOKE resonances. In particular, instead of one resonance at $\lambda$=850 nm for the plasmonic crystal, three resonances at $\lambda$=820 nm, $\lambda$=890 nm and $\lambda$=950 nm appear. As a result the structure acquires pronounced broadband magneto-optical response in the wavelength range from $\lambda$=780 nm to $\lambda$=980 nm. It makes the proposed structure very promising for numerous nanophotonics applications including optical sensing, control of light, all-optical control of magnetization etc.

**Funding**. Russian Foundation for Basic Research (16-52-45061); the Russian Presidential Grant (MK-2047.2017.2); Department of Science and Technology (DST), India; the Ministry of Education and Science of the Russian Federation (3.7126.2017/8.9).

**Acknowledgment**. The work is supported by the Russian Foundation for Basic Research (project 16-52-45061), the Russian Presidential Grant MK-2047.2017.2, and Department of Science and Technology (DST), India. A.N.S., A.R.P. and V.N.B. acknowledge support by the Ministry of Education and Science of the Russian Federation [project number 3.7126.2017/8.9].